\documentclass[a4paper,onecolumn,11pt]{article}
\usepackage{amsfonts}
\usepackage{graphicx}
\usepackage{amsmath}

\input{tcilatex}
\begin{document}

\title{Beams Propagation Modelled by Bi-filters}
\author{B. Lacaze \\
Tesa 14/16 Port St-Etienne 31000 Toulouse France\\
e-mail address: bernard.lacaze@tesa.prd.fr}
\maketitle

\begin{abstract}
In acoustic, ultrasonic or electromagnetic propagation, crossed media are
often modelled by linear filters with complex gains in accordance with the
Beer-Lambert law. This paper addresses the problem of propagation in media
where polarization has to be taken into account. Because waves are now
bi-dimensional, an unique filter is not sufficient to represent the effects
of the medium. We propose a model which uses four linear invariant filters,
which allows to take into account exchanges between components of the field.
We call it bi-filter because it has two inputs and two outputs. Such a
circuit can be fitted to light devices like polarizers, rotators and
compensators and to propagation in free space. We give a generalization of
the Beer-Lambert law which can be reduced to the usual one in some cases and
which justifies the proposed model for propagation of electromagnic beams in
continuous media.

\textit{keywords: }linear filtering, polarization, Beer-Lambert law, random
processes.
\end{abstract}

\section{Introduction}

Propagation of acoustic or electromagnetic beams is often studied through
linear differential equations with coefficients which depend on the medium
characteristics. Because of the difficulty to estimate parameters and to
give well-fitted equations, the medium is often taken to be a linear
invariant filter (LIF) with given spectral gains. For instance, propagation
of ultrasonics through biological tissues is taken to be a filter with
complex gain exp$\left[ -\alpha \omega ^{\beta }\right] .$ In scanning,
values of parameters allow diagnosis \cite{Kuc}, \cite{Lewi}, \cite{Laca4}.
The same model is used for losses and dispersion of radiowaves in coaxial
cables, free space or fiber optics \cite{Laca2}, \cite{Laca6}.

We consider beams reduced to a trajectory in some medium. Beams measurements
are very often reduced to the values of quantities available at two points,
one at the origin and one at the end point (for instance the amplitude or
the power). Even if the beam is received in a antenna with a given area, the
information is often added in a coaxial cable or in a wave guide, and
reduced to a finite number of complex quantities. Moreover, comparisons
about values at the transmitter and at the receiver give insights about the
medium, and it is the aim of most devices used in practice (and this is true
for waves in any frequency band, optics, radio, acoustics or ultrasonics).
Actually, the main difference between sonics or ultrasonics waves and radio
or optical waves is the polarization of the last one. Then they are defined
by vectors instead of scalars. This paper highlights this difference.

We define a one-dimensional beam by an electric field $\overrightarrow{%
\mathbf{E}^{z}}$=$\left\{ \overrightarrow{E^{z}}\left( t\right) ,t\in 
\mathbb{R}\right\} $ at each point of a trajectory. $t$ stands for the time, 
$z$ is the coordinate on the trajectory and $\overrightarrow{E^{z}}\left(
t\right) $ is a vector orthogonal to the trajectory. In numerous situations,
it is sufficient to consider the amplitude of $\overrightarrow{E^{z}}\left(
t\right) ,$ which defines the power and the spectral content of the wave.
Then, it is sufficient to model the medium as a linear filtering, as in the
case of acoustic wave propagation. It is no longer the case when
polarization phenomena have to be taken into account. We have to show the
evolution of two components $E_{x}^{z}\left( t\right) ,E_{y}^{z}\left(
t\right) $ of the electric field which defines the wave. Because both
components can be linked, the behavior of one component at the time $t$ can
depend on the other component. In a linear model, this means that each
component is the addition of linear filtering of itself and of the other
component. Consequently, the medium has to be defined by a family of four
filters. Nevertheless, we will talk about a bi-filter because the operation
defines a linear way between two couples of processes, two inputs and two
outputs. We place ourselves in a stationary frame where the processes are
stationary (in the wide sense) with stationary correlations. The filters
which model the medium will be time invariant with the acronym LIF\ for
Linear (Time) Invariant Filters and will be defined by complex gains rather
than impulse responses. They generalize the notion of \textquotedblleft
scattering matrix\textquotedblright\ used for monochromatic waves where each
coordinate of the scattered wave is a linear combination of transmitted
coordinates. Though power spectra of electromagnetic waves are the most
often assumed in limited bands, the definitions and computations will be
done for general spectra possibly with infinite support.

Section 1 of this paper addresses the definition and properties of
bi-filters. Sections 2 and 3 apply results of section 1 to one-dimensional
beams. Section 4 studies a generalization of the Beer-Lambert law which
shows why a continuous medium can be modelled by a bi-filter. Appendix gives
proofs of formulae.

\section{Bi-filtering}

1) We consider two (real or complex) stationary processes 
\begin{equation*}
\mathbf{X}_{1}=\left\{ X_{1}\left( t\right) ,t\in \mathbb{R}\right\} ,%
\mathbf{Y}_{1}=\left\{ Y_{1}\left( t\right) ,t\in \mathbb{R}\right\}
\end{equation*}
with spectral density $s_{X_{1}},s_{Y_{1}}$ depending on the frequency $%
\omega /2\pi ,$ but we will omit the variable $\omega $ when the result is
not ambiguous. Moreover we assume a stationary correlation between them
which defines a cross-spectrum $s_{X_{1}Y_{1}}$ such that \cite{Cram}, \cite%
{Papo} 
\begin{equation}
\left\{ 
\begin{array}{l}
\text{E}\left[ X_{1}\left( t\right) X_{1}^{\ast }\left( t-\tau \right) %
\right] =\int_{-\infty }^{\infty }s_{X_{1}}\left( \omega \right) e^{i\omega
\tau }d\omega \\ 
\text{E}\left[ Y_{1}\left( t\right) Y_{1}^{\ast }\left( t-\tau \right) %
\right] =\int_{-\infty }^{\infty }s_{Y_{1}}\left( \omega \right) e^{i\omega
\tau }d\omega \\ 
\text{E}\left[ X_{1}\left( t\right) Y_{1}^{\ast }\left( t-\tau \right) %
\right] =\int_{-\infty }^{\infty }s_{X_{1}Y_{1}}\left( \omega \right)
e^{i\omega \tau }d\omega%
\end{array}
\right.
\end{equation}
where E$\left[ ..\right] $ stands for the mathematical expectation (or
ensemble mean) and the superscript $^{\ast }$ stands for the complex
conjugate.

Now, we define the processes \textbf{X}$_{2}$=$\left\{ X_{2}\left( t\right)
,t\in \mathbb{R}\right\} ,$\textbf{Y}$_{2}$=$\left\{ Y_{2}\left( t\right)
,t\in \mathbb{R}\right\} $ by 
\begin{equation}
\left\{ 
\begin{array}{c}
X_{2}\left( t\right) =X_{1}\ast h_{11}\left( t\right) +Y_{1}\ast
h_{12}\left( t\right) \\ 
Y_{2}\left( t\right) =X_{1}\ast h_{21}\left( t\right) +Y_{1}\ast
h_{22}\left( t\right)%
\end{array}
\right.
\end{equation}
where the $h_{jk}$ can be considered as impulse responses of four linear
invariant filters $\mathcal{H}_{jk}$ and $\left( .\ast .\right) $ stands for
the convolution product$.$ For example we have 
\begin{equation*}
X_{1}\ast h_{21}\left( t\right) =\int_{-\infty }^{\infty }X_{1}\left(
u\right) h_{21}\left( t-u\right) du
\end{equation*}
We will say that $\left( \mathbf{X}_{1},\mathbf{Y}_{1}\right) $ and $\left( 
\mathbf{X}_{2},\mathbf{Y}_{2}\right) $ are the input and the output of the
bi-filter $\mathcal{H}=\left\{ \mathcal{H}_{jk},j,k=1,2\right\} .$ We know
that it is more convenient to use spectral gains $H_{jk}\left( \omega
\right) $ rather than impulse responses. When impulse responses are
sufficiently regular we have together 
\begin{equation}
\left\{ 
\begin{array}{l}
h_{jk}\left( t\right) =\frac{1}{2\pi }\int_{-\infty }^{\infty }H_{jk}\left(
\omega \right) e^{i\omega t}d\omega \\ 
H_{jk}\left( \omega \right) =\int_{-\infty }^{\infty }h_{jk}\left( t\right)
e^{-i\omega t}dt%
\end{array}
\right.
\end{equation}
but the $h_{jk}$ are not always ordinary functions. The following writing of 
$\left( 2\right) $ is more general because the impulse responses do not
appear 
\begin{equation}
\left\{ 
\begin{array}{c}
X_{2}\left( t\right) =\mathcal{H}_{11}\left[ \mathbf{X}_{1}\right] \left(
t\right) +\mathcal{H}_{12}\left[ \mathbf{Y}_{1}\right] \left( t\right) \\ 
Y_{2}\left( t\right) =\mathcal{H}_{21}\left[ \mathbf{X}_{1}\right] \left(
t\right) +\mathcal{H}_{22}\left[ \mathbf{Y}_{1}\right] \left( t\right)%
\end{array}
\right.
\end{equation}
It is proved in appendix 1 that the bi-dimensional process $\left( \mathbf{X}%
_{2},\mathbf{Y}_{2}\right) $ is stationary with spectral characteristics
perfectly defined by formulae $\left( 34\right) $ which are simplified in 
\begin{equation}
\left\{ 
\begin{array}{l}
s_{X_{2}Y_{2}}=H_{12}H_{22}^{\ast }s_{Y_{1}}+H_{11}H_{21}^{\ast
}s_{X_{1}}+H_{12}H_{21}^{\ast }s_{Y_{1}X_{1}}+H_{11}H_{22}^{\ast
}s_{Y_{1}X_{1}}^{\ast } \\ 
s_{X_{2}}=\left\vert H_{12}\right\vert ^{2}s_{Y_{1}}+\left\vert
H_{11}\right\vert ^{2}s_{X_{1}}+2\mathcal{R}\left[ H_{12}H_{11}^{\ast
}s_{Y_{1}X_{1}}\right] \\ 
s_{Y_{2}}=\left\vert H_{21}\right\vert ^{2}s_{X_{1}}+\left\vert
H_{22}\right\vert ^{2}s_{Y_{1}}+2\mathcal{R}\left[ H_{22}H_{21}^{\ast
}s_{Y_{1}X_{1}}\right] \\ 
s_{X_{1}X_{2}}=H_{12}^{\ast }s_{Y_{1}X_{1}}^{\ast }+H_{11}^{\ast }s_{X_{1}}
\\ 
s_{X_{1}Y_{2}}=H_{22}^{\ast }s_{Y_{1}X_{1}}^{\ast }+H_{21}^{\ast }s_{X_{1}}
\\ 
s_{Y_{1}X_{2}}=H_{11}^{\ast }s_{Y_{1}X_{1}}+H_{12}^{\ast }s_{Y_{1}} \\ 
s_{Y_{1}Y_{2}}=H_{21}^{\ast }s_{Y_{1}X_{1}}+H_{22}^{\ast }s_{Y_{1}}%
\end{array}
\right.
\end{equation}
where $\mathcal{R}\left[ ..\right] $ stands for the real part, $s_{a}$ and $%
s_{ab}$ are for spectra or cross-spectra, $H_{jk}$ for filters complex
gains, and all terms depend on $\omega $. The figure 1 gives the scheme of
the bi-filter. When used in the matrix form \textbf{H}=$\left[ H_{ij}\right]
,$ the bi-filter is a \textquotedblleft scattering matrix\textquotedblright\
for each value of $\omega .$\FRAME{ftbpF}{4.2163in}{3.3782in}{0pt}{}{}{%
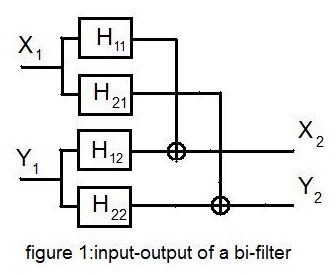}{\special{language "Scientific Word";type
"GRAPHIC";maintain-aspect-ratio TRUE;display "USEDEF";valid_file "F";width
4.2163in;height 3.3782in;depth 0pt;original-width 4.0258in;original-height
3.2203in;cropleft "0";croptop "1";cropright "1";cropbottom "0";filename
'C:/Users/Bernard/Desktop/waves/opt10-2.jpg';file-properties "XNPEU";}}

\section{Application to waves}

\subsection{General formulae}

1) We consider a beam which crosses a medium along the axis Oz of the
orthogonal trihedron Oxyz. The beam is defined by its electrical field $%
\overrightarrow{E^{z}}\left( t\right) =\left( E_{x}^{z}\left( t\right)
,E_{y}^{z}\left( t\right) \right) $ at time $t$ and distance $z,$ where
components (which are orthogonal to Oz) are taken on the axes Ox and Oy. We
assume

a) that the processes (at O) $\mathbf{E}_{x}^{0}=\left\{ E_{x}^{0}\left(
t\right) ,t\in \mathbb{R}\right\} $ and $\mathbf{E}_{y}^{0}=\left\{
E_{y}^{0}\left( t\right) ,t\in \mathbb{R}\right\} $ are stationary and with
stationary correlation. The spectra and the cross-spectrum are 
\begin{equation*}
s_{x}^{0}\left( \omega \right) ,s_{y}^{0}\left( \omega \right)
,s_{xy}^{0}\left( \omega \right)
\end{equation*}

b) and that the medium between the points $z_{1}$ and $z_{2}$ can be
characterized by four LIF (linear invariant filters) $\mathcal{H}%
_{jk}^{z_{1}z_{2}}$ with the complex gains $H_{jk}^{z_{1}z_{2}}\left( \omega
\right) $ such that, whatever $0\leq u\leq z$ 
\begin{equation}
\left\{ 
\begin{array}{l}
E_{x}^{z}\left( t\right) =\mathcal{H}_{11}^{uz}\left[ \mathbf{E}_{x}^{u}%
\right] \left( t\right) +\mathcal{H}_{12}^{uz}\left[ \mathbf{E}_{y}^{u}%
\right] \left( t\right) \\ 
E_{y}^{z}\left( t\right) =\mathcal{H}_{21}^{uz}\left[ \mathbf{E}_{x}^{u}%
\right] \left( t\right) +\mathcal{H}_{22}^{uz}\left[ \mathbf{E}_{y}^{u}%
\right] \left( t\right) .%
\end{array}
\right.
\end{equation}
This writing highlights the dependence of $E_{x}^{z}\left( t\right) $ and $%
E_{y}^{z}\left( t\right) $ on the whole processes $\mathbf{E}_{x}^{u}$ and $%
\mathbf{E}_{y}^{u}$ and not only on the r.v. $E_{x}^{u}\left( t\right) $ and 
$E_{y}^{u}\left( t\right) .$ Formulae $\left( 5\right) $ allow to write the
different spectra and cross-spectra of the beam as (with $H_{jk}=H_{jk}^{0z}$
to alleviate formulae$)$%
\begin{equation}
\left\{ 
\begin{array}{l}
s_{xy}^{z}=H_{12}H_{22}^{\ast }s_{y}^{0}+H_{11}H_{21}^{\ast
}s_{x}^{0}+H_{12}H_{21}^{\ast }s_{yx}^{0}+H_{11}H_{22}^{\ast }s_{xy}^{0} \\ 
s_{x}^{z}=\left\vert H_{12}\right\vert ^{2}s_{y}^{0}+\left\vert
H_{11}\right\vert ^{2}s_{x}^{0}+2\mathcal{R}\left[ H_{12}H_{11}^{\ast
}s_{yx}^{0}\right] \\ 
s_{y}^{z}=\left\vert H_{21}\right\vert ^{2}s_{x}^{0}+\left\vert
H_{22}\right\vert ^{2}s_{y}^{0}+2\mathcal{R}\left[ H_{22}H_{21}^{\ast
}s_{yx}^{0}\right] \\ 
s_{xx}^{0z}=H_{12}^{\ast }s_{xy}^{0}+H_{11}^{\ast }s_{x}^{0} \\ 
s_{xy}^{0z}=H_{22}^{\ast }s_{xy}^{0}+H_{21}^{\ast }s_{x}^{0} \\ 
s_{yx}^{0z}=H_{11}^{\ast }s_{yx}^{0}+H_{12}^{\ast }s_{y}^{0} \\ 
s_{yy}^{0z}=H_{21}^{\ast }s_{yx}^{0}+H_{22}^{\ast }s_{y}^{0}%
\end{array}
\right.
\end{equation}
where the seven equalities are respectively for the couples (from the top to
the bottom)

\begin{equation*}
\left( E_{x}^{z},E_{y}^{z}\right) ,\left( E_{x}^{z},E_{x}^{z}\right) ,\left(
E_{y}^{z},E_{y}^{z}\right) ,\left( E_{x}^{0},E_{x}^{z}\right) ,\left(
E_{x}^{0},E_{y}^{z}\right) ,\left( E_{y}^{0},E_{x}^{z}\right) ,\left(
E_{y}^{0},E_{y}^{z}\right) .
\end{equation*}
The power $P_{z}$ of the wave at $z$ is defined from the components by 
\begin{equation}
P_{z}=\text{E}\left[ \left\vert E_{x}^{z}\left( t\right) \right\vert
^{2}+\left\vert E_{y}^{z}\left( t\right) \right\vert ^{2}\right]
=\int_{-\infty }^{\infty }\left[ s_{x}^{z}+s_{y}^{z}\right] \left( \omega
\right) d\omega .
\end{equation}
This definition is justified (at a multiplicative constant in accordance
with some physical system of units) because devices for intensity
measurements are not sensitive to the direction of the electric field. Also,
the definition is independent of the basis used (see below).

2) In the proposed model, the filters complex gains are proper to the
coordinate system Oxy. For the system Ox'y' deduced by a rotation of angle $%
\theta ,$ the new system of complex gains $K_{jk}^{z_{1}z_{2}}$ is defined
by (we give the system for $K_{jk}^{0z}=K_{jk})$%
\begin{equation*}
\left[ 
\begin{array}{c}
K_{11} \\ 
K_{12} \\ 
K_{21} \\ 
K_{22}%
\end{array}
\right] =\mathbf{\Theta }\left[ 
\begin{array}{c}
H_{11} \\ 
H_{12} \\ 
H_{21} \\ 
H_{22}%
\end{array}
\right]
\end{equation*}
\begin{equation}
\Theta \mathbf{=}\left[ 
\begin{array}{cc}
\mathbf{P}\cos \theta & \mathbf{P}\sin \theta \\ 
-\mathbf{P}\sin \theta & \mathbf{P}\cos \theta%
\end{array}
\right] ,\mathbf{P=}\left[ 
\begin{array}{cc}
\cos \theta & \sin \theta \\ 
-\sin \theta & \cos \theta%
\end{array}
\right] .
\end{equation}
$\mathbf{\Theta }$ is orthogonal and then $\mathbf{\Theta }^{-1}=\mathbf{%
\Theta }^{t}$ (the transpose of $\mathbf{\Theta }$). Also, the only set of
parameters which are invariant by rotation verify (see appendix 2) 
\begin{equation}
H_{11}=H_{22},\text{ \ \ }H_{12}=-H_{21}.
\end{equation}
This set of $H_{jk}$ defines the proper subspace of the eigenvalue 1. This
property is very important and will often be used. Moreover it is easy to
verify that 
\begin{equation}
\left\{ 
\begin{array}{l}
H_{11}+H_{22}=K_{11}+K_{22} \\ 
H_{11}H_{22}-H_{12}H_{21}=K_{11}K_{22}-K_{12}K_{21}.%
\end{array}
\right.
\end{equation}
This property of invariance will be used in section 4.2.

3) The sum of bi-filters $\mathcal{K=H+H}^{\prime }$ is naturally defined by 
$K_{jk}=H_{jk}+H_{jk}^{\prime }$ whatever $j,k.$ It is the same as the usual
sum for linear filters (filters in parallel of the circuit theory). The
product of bi-filters $\mathcal{K=H}$x$\mathcal{H}^{\prime }$ is defined in
the same way as the filters in series of the circuit theory 
\begin{equation*}
\left( \mathbf{X}_{2},\mathbf{Y}_{2}\right) =\mathcal{K}\left[ \mathbf{X}%
_{1},\mathbf{Y}_{1}\right] =\mathcal{H}^{\prime }\left\{ \mathcal{H}\left[ 
\mathbf{X}_{1},\mathbf{Y}_{1}\right] \right\}
\end{equation*}
The scheme of this operation is given in figure 2 with the notations used in
the next sections. Equivalently, we have (because the complex gains of LIF
in parallel and LIF in series are the sum and the product of individual
complex gains) 
\begin{equation*}
\left\{ 
\begin{array}{c}
K_{11}=H_{11}H_{11}^{\prime }+H_{21}H_{12}^{\prime } \\ 
K_{12}=H_{12}H_{11}^{\prime }+H_{22}H_{12}^{\prime } \\ 
K_{21}=H_{11}H_{21}^{\prime }+H_{21}H_{22}^{\prime } \\ 
K_{22}=H_{12}H_{21}^{\prime }+H_{22}H_{22}^{\prime }.%
\end{array}
\right.
\end{equation*}
This operation is associative but not commutative (which is a huge
difference with ordinary LIF). The filter $\mathcal{I}$ such that 
\begin{equation*}
I_{11}=I_{22}=1,I_{12}=I_{21}=0
\end{equation*}
is the unit: $\mathcal{I}$x$\mathcal{H}=\mathcal{H}$x$\mathcal{I=H}$.

Sums of bi-filters have to be used when the beam is split so that the
results cross different media. Each of them corresponds to a bi-filter and
outputs are added (similar to a fringes pattern). The product is for a beam
which crosses two successive media, or two successive thickness of the same
medium, each of them being defined as a bi-filter.

\subsection{Polarized beam}

A \textquotedblleft polarized wave\textquotedblright\ at $z$ is defined by a
direction $\psi $ (with respect to the axis Ox), and some stationary process 
\textbf{A}$^{z}$\textbf{=}$\left\{ A^{z}\left( t\right) ,t\in \mathbb{R}%
\right\} $%
\begin{equation}
\left\{ 
\begin{array}{c}
E_{x}^{z}\left( t\right) =A^{z}\left( t\right) \cos \psi \\ 
E_{y}^{z}\left( t\right) =A^{z}\left( t\right) \sin \psi .%
\end{array}
\right.
\end{equation}
This definition is consistent because, in the basis Ox'y' such as $\left( 
\text{Ox, Ox'}\right) =\psi ^{\prime }$ we have 
\begin{equation*}
\left\{ 
\begin{array}{c}
E_{x^{\prime }}^{z}\left( t\right) =A^{z}\left( t\right) \cos (\psi -\psi
^{\prime }) \\ 
E_{y^{\prime }}^{z}\left( t\right) =A^{z}\left( t\right) \sin (\psi -\psi
^{\prime }).%
\end{array}
\right.
\end{equation*}
In this paper the \textquotedblleft polarized wave\textquotedblright\ is for
the \textquotedblleft linear polarized wave\textquotedblright\ used for
deterministic beams. Because the sum of two polarized waves $\left( \mathbf{A%
}^{z},\psi \right) $ and $\left( \mathbf{B}^{z},\phi \right) $ is generally
not polarized, the set of polarized waves has not an interesting algebraic
structure. But, such waves can be treated separately when transformations
are linear. It is worth-noting that the wave 
\begin{equation*}
\mathcal{F}\left[ \mathbf{E}_{x}^{z}\right] ,\mathcal{F}\left[ \mathbf{E}%
_{y}^{z}\right]
\end{equation*}
where $\mathcal{F}$ is any LIF, is polarized in the same direction that $%
\left( \mathbf{E}_{x}^{z},\mathbf{E}_{y}^{z}\right) $.

From $\left( 11\right) ,$ a necessary condition for a polarized wave is 
\begin{equation}
\rho _{x}^{z}\rho _{y}^{z}=\left( \rho _{xy}^{z}\right) ^{2}
\end{equation}
where 
\begin{equation*}
\rho _{x}^{z}=\text{E}\left[ \left\vert E_{x}^{z}\left( t\right) \right\vert
^{2}\right] ,\rho _{y}^{z}=\text{E}\left[ \left\vert E_{y}^{z}\left(
t\right) \right\vert ^{2}\right] ,\rho _{xy}^{z}=\text{E}\left[
E_{x}^{z}\left( t\right) E_{y}^{z\ast }\left( t\right) \right]
\end{equation*}
(and not $\left\vert \rho _{xy}^{z}\right\vert ^{2}$ except for real
processes). Conversely, if $\left( 13\right) $ is verified, $\rho _{xy}^{z}$
is real (because $\rho _{x}^{z}\rho _{y}^{z}\geq 0)$, which leads to 
\begin{equation*}
E\left[ \left\vert E_{x}^{z}\left( t\right) -\lambda E_{y}^{z}\left(
t\right) \right\vert ^{2}\right] =\left\vert \sqrt{\rho _{x}^{z}}-\lambda
\varepsilon \sqrt{\rho _{y}^{z}}\right\vert ^{2}
\end{equation*}
where $\varepsilon =\pm 1$ and $\lambda \in \mathbb{C}$ ($\varepsilon =1$
when $\rho _{xy}\geq 0).$ Consequently, we have $E_{x}^{z}\left( t\right)
=\lambda E_{y}^{z}\left( t\right) $ with $\lambda =\varepsilon \sqrt{\rho
_{x}^{z}/\rho _{y}^{z}}$ (quantity which is real and independent of $t)$ and 
$\left( 13\right) $ becomes a sufficient condition for polarization. From $%
\left( 12\right) $ the condition $\left( 13\right) $ is true whatever the
system Oxy. Reciprocally, when verified in one particular system, it is
verified in the others. Also, the equality $\left( 13\right) $ is true for
correlations and then for (regular) spectra 
\begin{equation}
s_{x}^{z}s_{y}^{z}=\left( s_{xy}^{z}\right) ^{2}
\end{equation}
but this last equality is not sufficient for polarization. As an example 
\begin{equation*}
\left\{ 
\begin{array}{c}
E_{x}^{z}\left( t\right) =A^{z}\left( t\right) \cos \psi \\ 
E_{y}^{z}\left( t\right) =B^{z}\left( t\right) \sin \psi%
\end{array}
\right.
\end{equation*}
where $\mathbf{B}^{z}$ is the output of a LIF with input $\mathbf{A}^{z}$
and complex gain $\varepsilon \left( \omega \right) $ taking only the values 
$\pm 1.$ The condition $\left( 14\right) $ is fulfilled but the wave is not
generally polarized and can be split in the sum of two polarized waves in
the directions $\psi $ and $-\psi .$

\subsection{Unpolarized beam}

1) We define an \textquotedblleft unpolarized wave\textquotedblright\ $%
\overrightarrow{\mathbf{E}^{z}}$\ at $z$ by the condition (whatever the
systems Oxy, Ox'y') 
\begin{equation*}
s_{xy}^{z}=s_{x^{\prime }y^{\prime }}^{z}=0.
\end{equation*}
Equivalently (see appendix 3) 
\begin{equation}
\left\{ 
\begin{array}{l}
s_{x}^{z}=s_{y}^{z}=s_{x^{\prime }}^{z}=s_{y^{\prime }}^{z} \\ 
s_{xy}^{z}=s_{x^{\prime }y^{\prime }}^{z}=0.%
\end{array}
\right.
\end{equation}
The term \textquotedblleft unpolarized wave\textquotedblright\ is used in
this paper rather than the term \textquotedblleft circular polarized
wave\textquotedblright\ which we encounter for deterministic waves. The
subset of uncorrelated unpolarized waves is a group for the addition. But
the sum of two correlated unpolarized waves can be not unpolarized. For
example, $\left( \mathbf{E}_{x},\mathbf{E}_{y}\right) $ added to $\left( 
\mathbf{E}_{x},-\mathbf{E}_{y}\right) $ is polarized along Ox.

In \cite{Mand}, pp. 350, and in the optics communauty, the unpolarized light
(natural light) is defined by the equality $\rho _{xy}^{z}=0,$ whatever Oxy,
which implies $\rho _{x}^{z}=\rho _{y}^{z}$. This condition is much more
weak than the condition $\left( 15\right) .$ In the strong definition, an
unpolarized wave remains unpolarized after crossing of a compensator. It is
not true when using the weak definition.

2) $\overrightarrow{\mathbf{E}^{0}}$ is an unpolarized wave when its
components $\mathbf{E}_{x}^{0}$ and $\mathbf{E}_{y}^{0}$ are uncorrelated
whatever the system Oxy (strong definition). It is the case if and only if
the components are uncorrelated for a given system in which the spectra are
identical ($s_{xy}^{0}=0,s_{x}^{0}=s_{y}^{0})$. In this circumstance, $%
\left( 7\right) $ is reduced to 
\begin{equation*}
\left\{ 
\begin{array}{l}
s_{xy}^{z}=\left( H_{12}H_{22}^{\ast }+H_{11}H_{21}^{\ast }\right) s_{x}^{0}
\\ 
s_{x}^{z}=\left( \left\vert H_{12}\right\vert ^{2}+\left\vert
H_{11}\right\vert ^{2}\right) s_{x}^{0} \\ 
s_{y}^{z}=\left( \left\vert H_{21}\right\vert ^{2}+\left\vert
H_{22}\right\vert ^{2}\right) s_{x}^{0}%
\end{array}
\right.
\end{equation*}
Consequently the wave $\overrightarrow{\mathbf{E}^{z}}$ remains unpolarized
whatever $s_{x}^{0}$ if and only if 
\begin{equation*}
\left\{ 
\begin{array}{l}
H_{12}H_{22}^{\ast }+H_{11}H_{21}^{\ast }=0 \\ 
\left\vert H_{12}\right\vert ^{2}+\left\vert H_{11}\right\vert
^{2}=\left\vert H_{21}\right\vert ^{2}+\left\vert H_{22}\right\vert ^{2}.%
\end{array}
\right.
\end{equation*}
Equivalently it exists a real function $P\left( \omega \right) $ such that 
\begin{equation}
\left\{ 
\begin{array}{l}
H_{22}\left( \omega \right) =H_{11}^{\ast }\left( \omega \right) e^{iP\left(
\omega \right) } \\ 
H_{12}\left( \omega \right) =-H_{21}^{\ast }\left( \omega \right)
e^{iP\left( \omega \right) }.%
\end{array}
\right.
\end{equation}

\subsection{Partially polarized beam}

Any wave which is not polarized or not unpolarized is a partially polarized
wave. This is a definition which results in different classes following the
definition given to unpolarization (weak or strong sense). In the class of
beams defined in appendix 5, a partially polarized beam corresponds to a
probability law different of the uniform on $\left( 0,2\pi \right) $ (case
of an unpolarized wave) and not degenerate (polarized wave when degenerate).

The Stokes decomposition theorem is still a studied problem \cite{Wolf1}. It
states that a partially polarized beam is the sum of a polarized beam and of
an unpolarized beam. For stationary processes with any spectra, solutions of
the problem can be found according to the set where solutions are searched,
and according to definition of unpolarization \cite{Laca5}. It is possible
to construct waves which are polarized in more than one direction. Provided
that the operations are linear, we can split the beam in a convenient number
of polarized ones and we can study them separately.

\subsection{Examples of bi-filters}

We characterize the following elementary bi-filters by the matrix \textbf{H}
of the $H_{jk}\left( \omega \right) $ in Oxy or the matrix \textbf{K }of the%
\textbf{\ }$K_{jk}\left( \omega \right) $ in Ox'y': 
\begin{equation*}
\mathbf{H=}\left[ 
\begin{array}{cc}
H_{11} & H_{12} \\ 
H_{21} & H_{22}%
\end{array}
\right]
\end{equation*}
The elementary operations which follow are often assumed independent of the
frequency. Actually this property is generally approached in the (limited)
frequency band of an experiment. However the matrix \textbf{H }which defines
the bi-filter is a function of the frequency $\omega /2\pi $ in most cases.

1) A compensator $\left( \theta _{x},\theta _{y}\right) $ between O and $z$
induces different delays $\theta _{x}$ and $\theta _{y}$ for the components: 
$E_{x}^{z}\left( t\right) =E_{x}^{0}\left( t-\theta _{x}\right) $ and $%
E_{y}^{z}=E_{y}^{0}\left( t-\theta _{y}\right) $ \cite{Wolf}$.$ The
equivalent bi-filter is defined by 
\begin{equation*}
\mathbf{H=}\left[ 
\begin{array}{cc}
e^{-i\omega \theta _{x}} & 0 \\ 
0 & e^{-i\omega \theta _{y}}%
\end{array}
\right]
\end{equation*}
In the basis Ox'y' defined by $\psi =$(Ox, Ox'), we have 
\begin{equation*}
\mathbf{K=}\left[ 
\begin{array}{cc}
e^{-i\omega \theta _{x}}\cos \psi & e^{-i\omega \theta _{y}}\sin \psi \\ 
-e^{-i\omega \theta _{x}}\sin \psi & e^{-i\omega \theta _{y}}\cos \psi%
\end{array}
\right]
\end{equation*}

2) The polarizer along the axis Ox' suppresses the orthogonal component
along Oy'. This defines the bi-filter (in the basis Oxy with $\psi =\left( 
\text{Ox,Ox'}\right) )$%
\begin{equation*}
\mathbf{H=}\left[ 
\begin{array}{cc}
\cos ^{2}\psi & \sin \psi \cos \psi \\ 
\sin \psi \cos \psi & \sin ^{2}\psi%
\end{array}
\right]
\end{equation*}

3) The compensator $\left( \theta _{x},\theta _{y}\right) $ followed by a
polarizer in the direction Ox' defines the bi-filter 
\begin{equation*}
\mathbf{H=}\left[ 
\begin{array}{cc}
e^{-i\omega \theta _{x}}\cos ^{2}\psi & e^{-i\omega \theta _{y}}\sin \psi
\cos \psi \\ 
e^{-i\omega \theta _{x}}\sin \psi \cos \psi & e^{-i\omega \theta _{y}}\sin
^{2}\psi%
\end{array}
\right]
\end{equation*}
The power $P^{z}$ at $z$ is given by, from $\left( 7\right) $ and $\left(
8\right) $%
\begin{equation*}
\left\{ 
\begin{array}{l}
P^{z}=\int_{-\infty }^{\infty }\left[ \alpha s_{x}^{0}+\beta s_{y}^{0}+2%
\mathcal{R}\left( \gamma s_{yx}^{0}\right) \right] \left( \omega \right)
d\omega \\ 
\alpha =\cos ^{2}\psi ,\text{ \ }\beta =\sin ^{2}\psi \\ 
\gamma =e^{-i\omega \left( \theta _{y}-\theta _{x}\right) }\sin \psi \cos
\psi .%
\end{array}
\right.
\end{equation*}
$P^{z}$ is a function of $\theta _{y}-\theta _{x}$ and $\psi .$ A good
choice of these parameters allows to measure particular values of $P^{z}.$
If the parameters do not depend on the frequency, this allows to estimate
the quantities 
\begin{equation*}
\left\{ 
\begin{array}{l}
\text{E}\left[ \left\vert E_{x}^{0}\left( t\right) \right\vert ^{2}\right]
=\int_{-\infty }^{\infty }s_{x}^{0}\left( \omega \right) d\omega \\ 
\text{E}\left[ \left\vert E_{y}^{0}\left( t\right) \right\vert ^{2}\right]
=\int_{-\infty }^{\infty }s_{y}^{0}\left( \omega \right) d\omega \\ 
\text{E}\left[ E_{x}^{0}\left( t\right) E_{y}^{0\ast }\left( t\right) \right]
=\int_{-\infty }^{\infty }s_{xy}^{0}\left( \omega \right) d\omega%
\end{array}
\right.
\end{equation*}
which define the Stokes parameters and others \cite{Mand}.

\subsection{Rotator}

In appendix 2, we prove that any polarized beam in the direction $\psi $
(with respect to Oxy) at O will be polarized in the direction $\psi +\theta $
at $z$ ($\theta \neq \frac{\pi }{2}\func{mod}\pi )$ if and only if 
\begin{equation}
\left\{ 
\begin{array}{l}
H_{21}\cos \theta =H_{11}\sin \theta \\ 
H_{12}=-H_{21} \\ 
H_{11}=H_{22}%
\end{array}
\right.
\end{equation}
The angle of polarization is changed by the quantity $\theta .$ Moreover,
the amplitudes at O and $z$ are linked through a LIF with complex gain $%
H_{11}/\cos \theta $%
\begin{equation}
\mathbf{A}^{z}=\frac{1}{\cos \theta }\mathcal{H}_{11}\left[ \mathbf{A}^{0}%
\right] .
\end{equation}
Note that the relations $\left( 17\right) $ and $\left( 18\right) $ are true
whatever the axes, because using $\left( 8\right) $ and $\left( 9\right) ,$
they imply $K_{jk}=H_{jk}$. If we want that the amplitude remains unchanged,
it is necessary that 
\begin{equation*}
H_{11}=\cos \theta
\end{equation*}
Here, the \textquotedblleft amplitude\textquotedblright\ is the coordinate
along the axis of polarization, and it is a quantity which can be complex.

More generally, any beam is defined by two polarized waves, the first one in
the direction Ox, the second one in the direction Oy. Each of them is
submitted to a rotation of angle $\theta .$ The amplitude of each of them is
the result of a LIF with complex gain $H_{11}/\cos \theta $ (which is a
quantity invariant in a change of basis). Then we can speak about a rotation
whatever the kind of wave, polarized or not.

Also we see that an unpolarized wave will be changed by the bi-filter in an
unpolarized wave such that (when $\theta \neq \frac{\pi }{2}\func{mod}\pi )$%
\begin{equation*}
s_{x}^{z}=s_{y}^{z}=\left\vert H_{11}\right\vert ^{2}\frac{s_{x}^{0}}{\cos
^{2}\theta }.
\end{equation*}
But more general conditions allow to retain the unpolarization.

\subsection{Depolarization}

The medium in free space or in optical fibers is a source of depolarization.
Bi-filters take into account this situation. If we start from a polarized
wave along Ox, we have by definition 
\begin{equation*}
\left\{ 
\begin{array}{c}
E_{x}^{z}\left( t\right) =\mathcal{H}_{11}\left[ \mathbf{E}_{x}^{0}\right]
\left( t\right) \\ 
E_{y}^{z}\left( t\right) =\mathcal{H}_{21}\left[ \mathbf{E}_{x}^{0}\right]
\left( t\right)%
\end{array}
\right.
\end{equation*}
Except when $\left[ H_{11}/H_{21}\right] \left( \omega \right) $ is a real
constant, the wave is no longer polarized (assuming $H_{11}$ and $H_{21}$
different of 0). With \ $\rho _{x}^{z},\rho _{y}^{z},\rho _{xy}^{z}$ as
defined in section 3.2, we define the constants $a,b,c,d$ by 
\begin{equation*}
\left\{ 
\begin{array}{c}
c=1-a=e^{i\nu }\rho _{x}^{z}/\rho ^{\prime } \\ 
d=-b=-e^{i\nu }\rho _{xy}^{z}/\rho ^{\prime } \\ 
\rho ^{\prime }=\sqrt{\rho _{x}^{z}\rho _{y}^{z}-\left\vert \rho
_{xy}^{z}\right\vert ^{2}}%
\end{array}
\right.
\end{equation*}
assuming that $\rho ^{\prime }\neq 0$ and where $\nu $ is any real number$.$
This leads to the Stokes decomposition \cite{Wolf1} 
\begin{equation*}
\left\{ 
\begin{array}{l}
E_{x}^{z}\left( t\right) =A\left( t\right) +B\left( t\right) \\ 
A\left( t\right) =aE_{x}^{z}\left( t\right) +bE_{y}^{z}\left( t\right) \\ 
B\left( t\right) =cE_{x}^{z}\left( t\right) +dE_{y}^{z}\left( t\right)%
\end{array}
\right.
\end{equation*}
with a polarized part $\left( \mathbf{A},\mathbf{0}\right) $ and an
unpolarized part (in the weak sense) $\left( \mathbf{B,E}_{y}^{z}\right) ,$
which verify the conditions 
\begin{equation*}
\text{E}\left[ B\left( t\right) E_{y}^{z\ast }\left( t\right) \right] =0,%
\text{ \ E}\left[ \left\vert B\left( t\right) \right\vert ^{2}\right] =\text{%
E}\left[ \left\vert E_{y}^{z}\left( t\right) \right\vert ^{2}\right] .
\end{equation*}
To perform this decomposition with a given power for each part, it suffices
to choose filters with complex gains $H_{11},H_{21}$ such that 
\begin{equation*}
\left\{ 
\begin{array}{l}
\rho _{y}^{z}=\int_{-\infty }^{\infty }\left[ \left\vert H_{21}\right\vert
^{2}s_{x}^{0}\right] \left( \omega \right) d\omega \\ 
\rho _{x}^{z}=\int_{-\infty }^{\infty }\left[ \left\vert H_{11}\right\vert
^{2}s_{x}^{0}\right] \left( \omega \right) d\omega \\ 
\rho _{xy}^{z}=\int_{-\infty }^{\infty }\left[ H_{11}H_{21}^{\ast }s_{x}^{0}%
\right] \left( \omega \right) d\omega .%
\end{array}
\right.
\end{equation*}
The condition $\rho ^{\prime }\neq 0$ can always be verified using
well-chosen filters. But other Stokes decompositions can be done, changing
the definition or the basis \cite{Laca5}.

\subsection{A particular class of beams}

The main problem is to know if the model is well-fitted to physical
situations. A particular model is treated in appendix 5. A
quasi-monochromatic light can be polarized, or unpolarized or partially
polarized. In the first case, the electric field has a constant direction
and in the second case, the electric field takes any direction with equal
probability (independently with its amplitude). They are extreme situations
and the intermediary situation is for a partially polarized light where the
field direction is a random process with a one-dimensional probability law
which is not degenerate (such as polarized light) neither uniformly
distributed on $\left( 0,2\pi \right) $ (such as unpolarized light)$.$ A
large class of light spectra verifies (see appendix 5) 
\begin{equation}
\left\{ 
\begin{array}{l}
s_{x}^{0}\left( \omega \right) =\alpha \left\vert f\left( \omega \right)
\right\vert ^{2} \\ 
s_{y}^{0}\left( \omega \right) =\left( 1-\alpha \right) \left\vert f\left(
\omega \right) \right\vert ^{2} \\ 
s_{xy}^{0}\left( \omega \right) =\beta \left\vert f\left( \omega \right)
\right\vert ^{2}%
\end{array}
\right.
\end{equation}
where $0\leq \alpha \leq 1,$ and $\left\vert \beta \right\vert \leq \sqrt{%
\alpha \left( 1-\alpha \right) }.$ The values $\left\vert \beta \right\vert =%
\sqrt{\alpha \left( 1-\alpha \right) }$ are for\ polarized light in a
direction defined by $\alpha ,$ and $\alpha =\frac{1}{2},\beta =0$ for
unpolarized light. \ A quasi-monochromatic wave corresponds to $f\left(
\omega \right) =0$ outside a short interval centered at some frequency $%
\omega _{0}/2\pi .$ $\left( 7\right) $ shows that the bi-filter does not
change the membership to the class if the parameters $H_{jk}$ are constant
on the spectral support of the wave. For instance, the wave $\left(
19\right) $ with parameters $\left( \alpha ,\beta ,f\right) $ is transformed
in the wave with parameters $\left( \alpha ^{\prime },\beta ^{\prime
},g\right) $ by the bi-filter 
\begin{equation}
\left\{ 
\begin{array}{c}
H_{22}=kH_{12} \\ 
H_{21}=kH_{11}%
\end{array}
\right.
\end{equation}
where the new parameters are ($k,\alpha ,\beta $ are scalar, $f$ and the $%
H_{jk}$ are functions of $\omega $) 
\begin{equation*}
\left\{ 
\begin{array}{l}
\alpha ^{\prime }=\left( 1+\left\vert k\right\vert ^{2}\right) ^{-1},\text{
\ \ \ \ \ }\beta ^{\prime }=k^{\ast }\left( 1+\left\vert k\right\vert
^{2}\right) ^{-1} \\ 
\left\vert g\right\vert ^{2}=\left\vert f\right\vert ^{2}\left( 1+\left\vert
k\right\vert ^{2}\right) \left( \left( 1-\alpha \right) \left\vert
H_{12}\right\vert ^{2}+\alpha \left\vert H_{11}\right\vert ^{2}+2\beta 
\mathcal{R}\left[ H_{12}H_{11}^{\ast }\right] \right) .%
\end{array}
\right.
\end{equation*}

\section{About the Beer-Lambert law}

\subsection{The Beer-Lambert law for filters}

In acoustics or ultrasonics, a beam is represented by a (real or complex)
scalar quantity $U^{z}\left( t\right) $ parametrized by the distance $z$ at
a transmitter. Very often, the propagation medium is modelled by a LIF. If
the medium can be split in \textquotedblleft independent\textquotedblright\
pieces, the complex gain of the equivalent filter verifies the functional
equation 
\begin{equation}
H^{z_{1}z_{2}}\left( \omega \right) =H^{z_{1}u}\left( \omega \right)
H^{uz_{2}}\left( \omega \right)
\end{equation}
where $H^{z_{1}z_{2}}\left( \omega \right) $ is the complex gain of the
piece of medium in the interval $\left( z_{1},z_{2}\right) .$ $\left(
21\right) $ expresses that the pieces $\left( z_{1},u\right) $ and $\left(
u,z_{2}\right) $ have the behavior of filters in series. Regular solutions
of $\left( 21\right) $ on $\mathbb{R}^{+}$ are in the form \cite{Laca3} 
\begin{equation*}
H^{uv}\left( \omega \right) =e^{a\left( v,\omega \right) -a\left( u,\omega
\right) }.
\end{equation*}
But if we suppose that 
\begin{equation*}
H^{uv}\left( \omega \right) =H^{v-u}\left( \omega \right)
\end{equation*}
(this means that only the thickness of the pieces appears) the last equality
is reduced to 
\begin{equation}
H^{z}\left( \omega \right) =e^{-za\left( \omega \right) }
\end{equation}
where $a\left( \omega \right) $ depends only on the medium properties. For
example, the model $a\left( \omega \right) =-\alpha \omega ^{2}$ ($\alpha
>0) $ is admitted in limited frequency bands for atmosphere or water
acoustic propagation, but other functions can be used \cite{Ette}, \cite%
{Park}, \cite{Szab}, \cite{Laca5}. For optical propagation in free space,
the function $a\left( \omega \right) $ is very complicated due to deep
absorption holes. $\left( 22\right) $ is the Beer-Lambert law used in many
domains of science. Finally, the spectral density $s^{z}\left( \omega
\right) $ and the power $P_{z}$ of the process \textbf{U}$^{z}$ verify 
\begin{equation*}
s^{z}\left( \omega \right) =e^{-2z\mathcal{R}\left[ a\left( \omega \right) %
\right] }s^{0}\left( \omega \right) ,\text{ \ \ \ }P_{z}=\int_{-\infty
}^{\infty }e^{-2z\mathcal{R}\left[ a\left( \omega \right) \right]
}s^{0}\left( \omega \right) d\omega .
\end{equation*}

\subsection{The Beer-Lambert law for bi-filters}

1) We ask the question to know if a generalization of the Beer-Lambert law
can be done when bi-filters are used. Of course, we assume that the
successive pieces of the medium do not interact and that their properties
are independent of the beams which cross them. Each piece in the interval $%
\left( u,v\right) $ is represented by four LIF of complex gains $H_{jk}^{uv}$
with respect to the basis Oxy. We assume some geometric coherence of the
medium so that 
\begin{equation}
H_{jk}^{uv}=H_{jk}^{v-u}.
\end{equation}
Figure 2 gives the equivalent circuit for two successive layers. \FRAME{ftbpF%
}{4.9917in}{5.051in}{0pt}{}{}{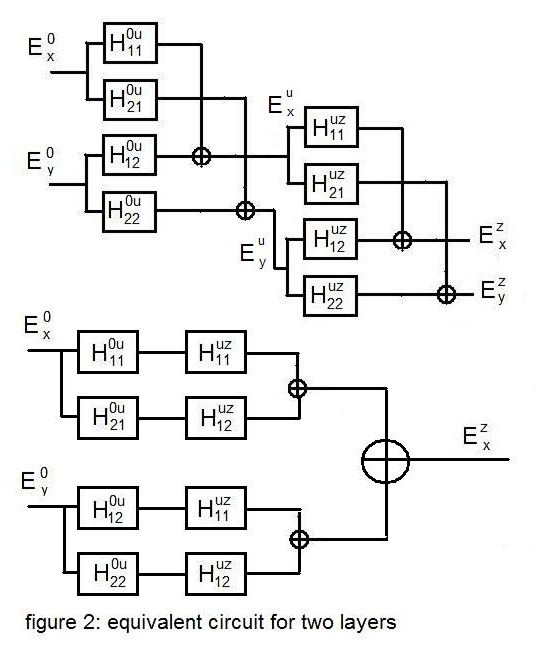}{\special{language "Scientific
Word";type "GRAPHIC";display "USEDEF";valid_file "F";width 4.9917in;height
5.051in;depth 0pt;original-width 7.2769in;original-height 7.287in;cropleft
"0";croptop "1";cropright "1";cropbottom "0";filename
'C:/Users/Bernard/Desktop/waves/opt10-4.jpg';file-properties "XNPEU";}}To
alleviate the formulae, we generally omit the variable $\omega .$
Consequently to the (strong) condition $\left( 23\right) $, and using the
scheme of figure 2, the problem can be translated in a set of differential
linear equations. We assume the existence of the derivatives $h_{jk}^{z}=%
\frac{\partial }{\partial z}H_{jk}^{z}$ for $z=0.$ We prove in appendix 4
(case 1) that, when $\lambda _{1},\lambda _{2}$ are distinct different of 0
roots of 
\begin{equation}
\lambda ^{2}-\left( h_{11}^{0}+h_{22}^{0}\right) \lambda
+h_{11}^{0}h_{22}^{0}-h_{12}^{0}h_{21}^{0}=0
\end{equation}
the only one solution of the problem is 
\begin{equation}
\left\{ 
\begin{array}{l}
H_{11}^{z}=\frac{h_{11}^{0}-\lambda _{2}}{\lambda _{1}-\lambda _{2}}%
e^{z\lambda _{1}}+\frac{\lambda _{1}-h_{11}^{0}}{\lambda _{1}-\lambda _{2}}%
e^{z\lambda _{2}} \\ 
H_{12}^{z}=\frac{h_{12}^{0}}{\lambda _{1}-\lambda _{2}}\left( e^{z\lambda
_{1}}-e^{z\lambda _{2}}\right) \\ 
H_{21}^{z}=\frac{h_{21}^{0}}{\lambda _{1}-\lambda _{2}}\left( e^{z\lambda
_{1}}-e^{z\lambda _{2}}\right) \\ 
H_{22}^{z}=\frac{h_{22}^{0}-\lambda _{2}}{\lambda _{1}-\lambda _{2}}%
e^{z\lambda _{1}}+\frac{\lambda _{1}-h_{22}^{0}}{\lambda _{1}-\lambda _{2}}%
e^{z\lambda _{2}}.%
\end{array}
\right.
\end{equation}
These equations are for bi-filters the version of the Beer-Lambert law for
filters where the medium is represented by the family of complex gains $%
H^{z}\left( \omega \right) =e^{-za\left( \omega \right) }.$ Also equations $%
\left( 25\right) $ define the set of admissible representations of a
continuous medium with bi-filters in series. Firstly we study the case where
the derivatives $h_{jk}^{z}=\frac{\partial }{\partial z}H_{jk}^{z}$ verify
the conditions $\ $%
\begin{equation}
h_{11}^{0}=h_{22}^{0},\text{ \ }h_{12}^{0}=-h_{21}^{0}.
\end{equation}
2) Let assume that $h_{12}^{0}$ does not depend on the frequency $\omega $/$%
2\pi .$ We find that the complex gains $H_{jk}^{z}$ verify 
\begin{equation}
\left\{ 
\begin{array}{l}
H_{11}^{z}=H_{22}^{z},\text{ \ \ \ \ \ \ }H_{12}^{z}=-H_{21}^{z} \\ 
H_{21}^{z}=e^{zh_{11}^{0}}\sin zh_{12}^{0},\text{ \ \ \ }%
H_{11}^{z}=e^{zh_{11}^{0}}\cos zh_{12}^{0}.%
\end{array}
\right.
\end{equation}
If $h_{12}^{0}$ is a real quantity, we recognize the formulas for a rotation
of the polarized beam (sections 3.6 and 6.2)$.$ The angle of rotation is
equal to $-zh_{12}^{0}$ (proportional to $z),$ and the amplitude $\mathbf{A}%
^{z}$ is the result of a LIF with input $\mathbf{A}^{0}$ and complex gain $%
e^{zh_{11}^{0}}$ ($h_{11}^{0}$ can depend on $\omega ).$ Then the
Beer-Lambert law $\left( 22\right) $ is true for the amplitude when the
relations $\left( 26\right) $ are fulfilled for real $h_{12}^{0}.$ Also $%
h_{11}^{0}$ is not real, because the imaginary part holds the propagation
time between the points O and $z$ (which cannot cancel and which depends on
the frequency $\omega /2\pi $ in case of dispersion). The real part measures
the weakening of the wave between both points.

To summarize, we can choose the parameters of the bi-filter to model a beam
with a given polarization at 0, which is rotated by any angle, and with any
weakening and with any propagation time. The (generally complex) amplitude
obeys the usual Beer-Lambert law. The angle is ruled by $h_{12}^{0},$ the
weakening by $\mathcal{R}\left[ h_{11}^{0}\right] $, the phase by $\mathcal{I%
}\left[ h_{11}^{0}\right] $, and the three parameters are proportional to
the distance $z$.

3) Now we consider any beam with its components $E_{x}^{0}\left( t\right)
,E_{y}^{0}\left( t\right) .$ When $\left( 26\right) $ is verified for the $%
h_{jk}^{0}$, these equalities are still true for the $H_{jk}^{z},$ and these
quantities are invariant by rotation ($K_{jk}^{z}=H_{jk}^{z}$ using $\left(
9\right) $ and $\left( 10\right) $). Then both components have the same
behavior when crossing the medium, i.e. are rotated, delayed and weakened by
same quantities. In the basis Ox'y' such that $-zh_{12}^{0}=\left( \text{Ox,
Ox'}\right) ,$ we have 
\begin{equation*}
\begin{array}{c}
E_{x^{\prime }}^{z}\left( t\right) =\mathcal{F}\left[ \mathbf{E}_{x}^{0}%
\right] \left( t\right) ,\text{ \ }E_{y^{\prime }}^{z}\left( t\right) =%
\mathcal{F}\left[ \mathbf{E}_{y}^{0}\right] \left( t\right)  \\ 
F\left( \omega \right) =e^{zh_{11}^{0}\left( \omega \right) }%
\end{array}%
\end{equation*}%
where $F\left( \omega \right) $ is the complex gain of the LIF $\mathcal{F}.$
Consequently the power at $z$ will be given by (from $\left( 8\right) $ and
the Wiener-Lee relations, see appendix 1$)$%
\begin{equation}
P_{z}=\int_{-\infty }^{\infty }e^{2z\mathcal{R}\left[ h_{11}^{0}\left(
\omega \right) \right] }\left[ s_{x}^{0}+s_{y}^{0}\right] \left( \omega
\right) d\omega 
\end{equation}%
to be compared with the result in the standard case (see the end of section
4.1). Then the power $P_{z}$ at $z$ is independent of the rotation defined
by the parameter $h_{12}^{0}.$ Equivalently, when studying the power, the
beam state of polarization is irrelevant, and the medium can be viewed as a
simple LIF where the Beer-Lambert law is available with the parameter $%
h_{11}^{0}$.

However, if we leave off the hypothesis of independence of $h_{12}^{0}$ with
the frequency, the interpretation of the bi-filter as a rotator is not true
because the angle of rotation is now a quantity proper to each frequency$.$

4) When $h_{12}^{0}=\rho e^{i\phi }$ is no longer real (but still
independent of the frequency), $\left( 27\right) $ is true but with complex
trigonometric functions which are expanded in 
\begin{equation*}
\left\{ 
\begin{array}{c}
H_{11}^{z}=a\cos \left( z\rho \cos \phi \right) +b\cos \left( z\rho \cos
\phi -\frac{\pi }{2}\right)  \\ 
H_{12}^{z}=a\sin \left( z\rho \cos \phi \right) +b\sin \left( z\rho \cos
\phi -\frac{\pi }{2}\right)  \\ 
a=e^{zh_{11}^{0}}\cosh \left( z\rho \sin \phi \right)  \\ 
b=-ie^{zh_{11}^{0}}\sinh \left( z\rho \sin \phi \right) .%
\end{array}%
\right. 
\end{equation*}%
This means that a polarized wave is split in two parts, the first one
rotated by the angle $-z\rho $ cos$\phi ,$ with amplitude coming from a LIF $%
\mathcal{A}$ with complex gain $a,$ the second one rotated by $\left( -z\rho
cos\phi +\frac{\pi }{2}\right) $ and with the complex gain $b$ of a LIF $%
\mathcal{B}$. The power $P_{z}$ is given by (the calculus is performed in
Ox'y' with $-z\rho cos\phi =$(Ox, Ox')) 
\begin{equation*}
P_{z}=\text{E}\left\{ \left[ \left\vert \mathcal{A}\left[ \mathbf{E}_{x}^{0}%
\right] -\mathcal{B}\left[ \mathbf{E}_{y}^{0}\right] \right\vert
^{2}+\left\vert \mathcal{B}\left[ \mathbf{E}_{x}^{0}\right] +\mathcal{A}%
\left[ \mathbf{E}_{y}^{0}\right] \right\vert ^{2}\right] \left( t\right)
\right\} 
\end{equation*}%
Computations can be performed (to simplify we give results for an
unpolarized wave $s_{xy}^{0}=0)$%
\begin{equation*}
P_{z}=\int_{-\infty }^{\infty }\left[ e^{2z\mathcal{R}\left[ h_{11}^{0}%
\right] }\left( s_{x}^{0}+s_{y}^{0}\right) \cosh \left( 2\rho z\sin \phi
\right) \right] \left( \omega \right) d\omega .
\end{equation*}%
We see that $\left( 28\right) $ is no longer true. 

Assume that the wave is polarized at O. Two reasons lead to a depolarization
of the wave. The first one is the dependency of the parameter $h_{12}^{0}$
on frequency. Then the angle of rotation of the beam is different following
the frequency. The second one happens when $h_{12}^{0}$ is not real, which
creates a secondary wave with amplitude $b$ and orthogonal to the main wave
(of amplitude $a)$. Because the function tanh is increasing on $\mathbb{R}$,
we have the same property for the quotient $\left\vert b/a\right\vert $ (but
which cannot reach 1)$.$

5) We go back to the general formulas $\left( 25\right) .$ They show that
the beam can be split in two parts, the first one containing the terms with
coefficient $e^{z\lambda _{1}}$ and the second part with $\ e^{z\lambda
_{2}}.$ Generally $\lambda _{1}$ and $\lambda _{2}$ are not conjugate
complex numbers because the equation $\left( 24\right) $ can have complex
coefficients. Assume that the input is the pure wave $e^{i\omega _{0}t}$. If 
\begin{equation*}
\lambda _{1}=a+ib,\text{ \ \ }\lambda _{2}=c+id
\end{equation*}%
the wave at $z$ will be the sum of two pure waves which cross the medium
with celerities $-\omega _{0}/b$ for the first one and $-\omega _{0}/d$\ for
the second one and with attenuation ruled by $a$ and $c$. Celerities are
different when $b\neq d$ as in a birefringent medium. Obviously, each wave
obeys the Beer-Lambert law in its simplest form but not the sum, except when
the $h_{jk}$ verify some conditions. Finally, equations $\left( 25\right) $
are given when the roots of $\left( 24\right) $ are distinct and different
of 0. Other situations are described in the appendix 4.

\section{Conclusion}

A bi-filter is defined by a circuit of four linear invariant filters. It is
a particular case 2x2 of the MIMO circuits (for multiple inputs-multiple
outputs) which are used for instance in communications between systems of
antennas \cite{Paul} and in sampling to improve the reconstruction of
signals \cite{Venk}. It generalizes the well-known notion of
\textquotedblleft scattering matrix\textquotedblright\ in radar processing.
In this paper, we address the problem of modelling a medium crossed by a
beam with two components and with some degree of polarization. We assume
that the beams are random processes with properties of stationarity and
spectra with any bandwith. We prove that bi-filters explain elementary
operations on electromagnetic waves and we establish a generalization of the
Beer-Lambert law which justifies the model in continuous media. Obviously,
other situations can be highlighted. The parameters of the bi-filter are
defined by the physical properties of the medium. Perhaps theoretical
considerations about the crossed material could allow the determination of
these parameters but I feel that it is a difficult task. In the field of
ultrasonics, for instance, the attenuation and the celerity of waves in some
medium (sea water, biological tissues...) are obtained by experiments and
not from the mechanical, chemical....considerations. I believe that the same
applies for electromagnetic waves. Characteristics of a material are
measured by studying a set of waves with different frequency and
polarization. These experiments are able to estimate the four complex gains
which define a bi-filter and they can help to give a fair representation of
the medium.

\section{Appendices}

\subsection{Appendix 1: Bi-filter}

1) Firstly, we summarize the usual theory of linear invariant filtering
(LIF) of stationary processes \cite{Cram}, \cite{Papo}, \cite{Laca1}. If $%
\mathbf{X}=\left\{ X\left( t\right) ,t\in \mathbb{R}\right\} $ is
characterized by its spectral density $s_{X}\left( \omega \right) ,$ it is
possible to define an isometry $I_{X}$ between the Hilbert spaces $\mathbf{H}%
_{X}$ and $\mathbf{K}_{s}$ where

-$\mathbf{H}_{X}$ is the set of linear combinations of the random variables $%
X\left( t\right) $ (completed by the closure of this set) when the scalar
product $\left\langle .,.\right\rangle _{H}$ and the associated distance $d$
are used: 
\begin{equation*}
\left\{ 
\begin{array}{l}
\left\langle X\left( u\right) ,X\left( v\right) \right\rangle _{\mathbf{H}}=%
\text{E}\left[ X\left( u\right) X^{\ast }\left( v\right) \right] \text{ \ \
\ } \\ 
d^{2}\left( A,B\right) =\text{E}\left[ \left\vert A-B\right\vert ^{2}\right]%
\end{array}
\right.
\end{equation*}

-$\mathbf{K}_{s}$ is the set of the $f\left( \omega \right) $ (from $\mathbb{%
R}$ to $\mathbb{C)}$ such as $\int_{-\infty }^{\infty }\left\vert f\left(
\omega \right) \right\vert ^{2}s_{X}\left( \omega \right) d\omega <\infty ,$
with the scalar product $\left\langle .,.\right\rangle _{K}$ and distances
defined by 
\begin{equation*}
\left\{ 
\begin{array}{c}
\left\langle f,g\right\rangle _{\mathbf{K}}=\int_{-\infty }^{\infty }\left[
fg^{\ast }s_{X}\right] \left( \omega \right) d\omega \text{ \ \ \ }. \\ 
\text{\ }d^{2}\left( f,g\right) =\int_{-\infty }^{\infty }\left[ \left\vert
f-g\right\vert ^{2}s_{X}\right] \left( \omega \right) d\omega%
\end{array}
\right. \text{\ }
\end{equation*}
The isometry $I_{X}$ is defined by the relation 
\begin{equation}
X\left( t\right) \longleftrightarrow _{I_{X}}e^{i\omega t}
\end{equation}
The isometry maintains the scalar product and the distances of corresponding
elements of the spaces. Consequently, it allows to perform calculations in
the space \textbf{K}$_{s}$ rather than in \textbf{H}$_{X},$ using Fourier
analysis and geometry of Hilbert spaces.

If $\mathcal{F}$ and $\mathcal{G}$ are two LIF with complex gains $F\left(
\omega \right) $ and $G\left( \omega \right) ,$ input \textbf{X}, outputs 
\textbf{U}=$\mathcal{F}\left[ \mathbf{X}\right] $ and \textbf{V}=$\mathcal{G}%
\left[ \mathbf{X}\right] ,$ we have 
\begin{equation}
\left\{ 
\begin{array}{l}
U\left( t\right) \longleftrightarrow _{I_{X}}F\left( \omega \right)
e^{i\omega t},\text{ \ }V\left( t\right) \longleftrightarrow _{I_{X}}G\left(
\omega \right) e^{i\omega t} \\ 
\text{E}\left[ U\left( t\right) V^{\ast }\left( t-\tau \right) \right]
=\int_{-\infty }^{\infty }e^{i\omega \tau }\left[ FG^{\ast }s_{X}\right]
\left( \omega \right) d\omega \\ 
s_{U}\left( \omega \right) =\left[ \left\vert F\right\vert ^{2}s_{X}\right]
\left( \omega \right) ,\text{ \ \ }s_{V}\left( \omega \right) =\left[
\left\vert G\right\vert ^{2}s_{X}\right] \left( \omega \right) \\ 
s_{UV}\left( \omega \right) =\left[ FG^{\ast }s_{X}\right] \left( \omega
\right)%
\end{array}
\right.
\end{equation}
where $s_{U}\left( \omega \right) ,s_{V}\left( \omega \right) ,s_{UV}\left(
\omega \right) $ are spectral and cross-spectral densities.

Finally, if \textbf{W}=$\mathcal{H}\left[ \mathbf{U}\right] ,$ and if $%
H\left( \omega \right) $ is the complex gain of the LIF $\mathcal{H},$ we
have 
\begin{equation*}
W\left( t\right) \longleftrightarrow _{I_{X}}\left[ HF\right] \left( \omega
\right) e^{i\omega t}
\end{equation*}
which is the relation for filters in series.

2) In section 2, the processes $\mathbf{X}_{1}$ and $\mathbf{Y}_{1}$ are
stationary and stationary correlated. If we look at processes $\mathbf{Y}%
_{1}^{\prime }$ and $\mathbf{Y}_{1}^{\prime \prime }$ defined by 
\begin{equation}
\left\{ 
\begin{array}{l}
Y_{1}\left( t\right) =Y_{1}^{\prime }\left( t\right) +Y_{1}^{\prime \prime
}\left( t\right) \\ 
Y_{1}^{^{\prime }}\left( t\right) \longleftrightarrow _{I_{X_{1}}}\left[ 
\frac{s_{Y_{1}X_{1}}}{s_{X_{1}}}\right] \left( \omega \right) e^{i\omega t}%
\end{array}
\right.
\end{equation}
\textbf{X}$_{1}$ and $\mathbf{Y}_{1}^{^{\prime }}$ are the input and the
output of a LIF filter of complex gain $s_{Y_{1}X_{1}}/s_{X_{1}}.$ Using $%
\left( 30\right) $ we obtain, whatever $t,\tau \in \mathbb{R}$%
\begin{equation*}
\text{E}\left[ X_{1}\left( t\right) Y_{1}^{\prime \prime \ast }\left( t-\tau
\right) \right] =0.
\end{equation*}
This means that $Y_{1}^{\prime }\left( t\right) $ is the orthogonal
projection of $Y_{1}\left( t\right) $ on \textbf{H}$_{X_{1}}$ and that $%
Y_{1}^{\prime \prime }\left( t\right) $ is orthogonal to \textbf{H}$%
_{X_{1}}, $ and 
\begin{equation}
\left\{ 
\begin{array}{l}
Y_{1}^{\prime }\left( t\right) \in \mathbf{H}_{X_{1}}\text{ \ \ \ \ }%
Y_{1}^{\prime \prime }\left( t\right) \perp \mathbf{H}_{X_{1}} \\ 
s_{Y_{1}^{\prime }}=\frac{\left\vert s_{Y_{1}X_{1}}\right\vert ^{2}}{%
s_{X_{1}}},\text{ \ }s_{X_{1}Y_{1}^{\prime }}=s_{X_{1}Y_{1}} \\ 
s_{Y_{1}^{\prime \prime }}=s_{Y_{1}}-\frac{\left\vert
s_{Y_{1}X_{1}}\right\vert ^{2}}{s_{X_{1}}}%
\end{array}
\right.
\end{equation}
where $s_{Y_{1}^{\prime }},s_{X_{1}Y_{1}^{\prime }},s_{Y_{1}^{\prime \prime
}}...$ are spectral and cross-spectral densities. Consequently, $\left(
4\right) $ can be split in two orthogonal systems $\mathbf{S}^{\prime }$ and 
$\mathbf{S}^{\prime \prime }$ 
\begin{eqnarray}
&&\mathbf{S}^{\prime }\left\{ 
\begin{array}{c}
X_{2}^{\prime }\left( t\right) =\mathcal{H}_{11}\left[ \mathbf{X}_{1}\right]
\left( t\right) +\mathcal{H}_{12}\left[ \mathbf{Y}_{1}^{\prime }\right]
\left( t\right) \\ 
Y_{2}^{\prime }\left( t\right) =\mathcal{H}_{21}\left[ \mathbf{X}_{1}\right]
\left( t\right) +\mathcal{H}_{22}\left[ \mathbf{Y}_{1}^{\prime }\right]
\left( t\right)%
\end{array}
\right. \\
&&\mathbf{S}^{\prime \prime }\left\{ 
\begin{array}{c}
X_{2}^{\prime \prime }\left( t\right) =\mathcal{H}_{12}\left[ \mathbf{Y}%
_{1}^{\prime \prime }\right] \left( t\right) \\ 
Y_{2}^{\prime \prime }\left( t\right) =\mathcal{H}_{22}\left[ \mathbf{Y}%
_{1}^{\prime \prime }\right] \left( t\right)%
\end{array}
\right.  \notag \\
&& 
\begin{array}{c}
Y_{1}\left( t\right) =Y_{1}^{\prime }\left( t\right) +Y_{1}^{\prime \prime
}\left( t\right) \\ 
Y_{2}\left( t\right) =Y_{2}^{\prime }\left( t\right) +Y_{2}^{\prime \prime
}\left( t\right) \\ 
X_{2}\left( t\right) =X_{2}^{\prime }\left( t\right) +X_{2}^{\prime \prime
}\left( t\right)%
\end{array}
\notag
\end{eqnarray}
By construction, the sets $\mathbf{S}^{\prime }=\left( \mathbf{X}%
_{2}^{\prime },\mathbf{Y}_{2}^{\prime }\right) $ and $\mathbf{S}^{\prime
\prime }=\left( \mathbf{X}_{2}^{\prime \prime },\mathbf{Y}_{2}^{\prime
\prime }\right) $ are uncorrelated. Each equation of $\mathbf{S}^{\prime }$
is equivalent to a circuit composed by three filters. $\mathbf{X}%
_{2}^{\prime }$ is the output of the LIF with complex gain $H_{11}+\frac{%
s_{Y_{1}X_{1}}}{s_{X_{1}}}H_{12}$ with input $\mathbf{X}_{1}.$ The filter of
complex gain $H_{21}+\frac{s_{Y_{1}X_{1}}}{s_{X_{1}}}H_{22}$ is for $\mathbf{%
Y}_{2}^{\prime }.$

3) From $\left( 30\right) ,\left( 32\right) ,\left( 33\right) $ we deduce
the spectral characteristics of $\left( \mathbf{Y}_{1},\mathbf{Y}_{2}\right) 
$ (we omit the variable $\omega )$ 
\begin{equation}
\left\{ 
\begin{array}{l}
s_{X_{2}Y_{2}}=\left( \frac{s_{Y_{1}X_{1}}}{s_{X_{1}}}H_{12}+H_{11}\right)
\left( \frac{s_{Y_{1}X_{1}}}{s_{X_{1}}}H_{22}+H_{21}\right) ^{\ast
}s_{X_{1}}+H_{12}H_{22}^{\ast }\left[ s_{Y_{2}}-\frac{\left\vert
s_{Y_{1}X_{1}}\right\vert ^{2}}{s_{X_{1}}}\right] \\ 
s_{X_{2}}=\left\vert \frac{s_{Y_{1}X_{1}}}{s_{X_{1}}}H_{12}+H_{11}\right%
\vert ^{2}s_{X_{1}}+\left\vert H_{12}\right\vert ^{2}\left[ s_{Y_{1}}-\frac{%
\left\vert s_{Y_{1}X_{1}}\right\vert ^{2}}{s_{X_{1}}}\right] \\ 
s_{Y_{2}}=\left\vert \frac{s_{Y_{1}X_{1}}}{s_{X_{1}}}H_{22}+H_{21}\right%
\vert ^{2}s_{X_{1}}+\left\vert H_{22}\right\vert ^{2}\left[ s_{Y_{1}}-\frac{%
\left\vert s_{Y_{1}X_{1}}\right\vert ^{2}}{s_{X_{1}}}\right] \\ 
s_{X_{1}X_{2}}=\left( \frac{s_{Y_{1}X_{1}}}{s_{X_{1}}}H_{12}+H_{11}\right)
^{\ast }s_{X_{1}} \\ 
s_{X_{1}Y_{2}}=\left( \frac{s_{Y_{1}X_{1}}}{s_{X_{1}}}H_{22}+H_{21}\right)
^{\ast }s_{X_{1}} \\ 
s_{Y_{1}X_{2}}=\frac{s_{Y_{1}X_{1}}}{s_{X_{1}}}\left( \frac{s_{Y_{1}X_{1}}}{%
s_{X_{1}}}H_{12}+H_{11}\right) ^{\ast }s_{X_{1}}+H_{12}^{\ast }\left[
s_{Y_{1}}-\frac{\left\vert s_{Y_{1}X_{1}}\right\vert ^{2}}{s_{X_{1}}}\right]
\\ 
s_{Y_{1}Y_{2}}=\frac{s_{Y_{1}X_{1}}}{s_{X_{1}}}\left( \frac{s_{Y_{1}X_{1}}}{%
s_{X_{1}}}H_{22}+H_{21}\right) ^{\ast }s_{X_{1}}+H_{22}^{\ast }\left[
s_{Y_{1}}-\frac{\left\vert s_{Y_{1}X_{1}}\right\vert ^{2}}{s_{X_{1}}}\right]%
\end{array}
\right.
\end{equation}

\subsection{Appendix 2: Rotator}

1) We study bi-filters which have polarized beams as input and output. We
look for the bi-filters which increase the angle of polarization by a given
quantity $\theta ,$ independently of the absolute values of the angle at the
input and independently of the power spectra.

Assume that the beam is polarized at 0 and $z$ with angles $\psi $ and $\psi
^{\prime }=\psi +\theta $ (with respect to Oxy)$.$ It is equivalent to have
(section 3-1) 
\begin{equation*}
\left\{ 
\begin{array}{c}
A^{z}\left( t\right) \cos \psi ^{\prime }=\left[ \mathcal{H}_{11}\cos \psi +%
\mathcal{H}_{12}\sin \psi \right] \left[ \mathbf{A}^{0}\right] \left(
t\right) \\ 
A^{z}\left( t\right) \sin \psi ^{\prime }=\left[ \mathcal{H}_{21}\cos \psi +%
\mathcal{H}_{22}\sin \psi \right] \left[ \mathbf{A}^{0}\right] \left(
t\right)%
\end{array}
\right.
\end{equation*}
Taking $\psi =0$ and $\psi =\pi /2$ leads to 
\begin{equation*}
\left\{ 
\begin{array}{c}
\mathcal{H}_{21}\left[ \mathbf{A}^{0}\right] =\mathcal{H}_{11}\left[ \mathbf{%
A}^{0}\right] \tan \theta \\ 
\mathcal{H}_{22}\left[ \mathbf{A}^{0}\right] =-\mathcal{H}_{12}\left[ 
\mathbf{A}^{0}\right] \cot \theta%
\end{array}
\right.
\end{equation*}
equalities which are true for any $\mathbf{A}^{0}.$ Then the equalities
about the complex gains become 
\begin{equation}
\left\{ 
\begin{array}{c}
H_{21}\cos \theta =H_{11}\sin \theta \\ 
H_{12}\cos \theta =-H_{22}\sin \theta%
\end{array}
\right.
\end{equation}
We enter $\left( 35\right) $ in the first equality which becomes 
\begin{equation*}
\left\{ 
\begin{array}{c}
A^{z}\left( t\right) \cos \left( \psi +\theta \right) =\left[ \mathcal{H}%
_{11}\cos \psi -\mathcal{H}_{22}\sin \psi \tan \theta \right] \left[ \mathbf{%
A}^{0}\right] \left( t\right) \\ 
A^{z}\left( t\right) \sin \left( \psi +\theta \right) =\left[ \mathcal{H}%
_{11}\cos \psi \tan \theta +\mathcal{H}_{22}\sin \psi \right] \left[ \mathbf{%
A}^{0}\right] \left( t\right)%
\end{array}
\right.
\end{equation*}
and we deduce the equality 
\begin{equation*}
\cot \left( \psi +\theta \right) =\frac{H_{11}\cos \psi \cos \theta
-H_{22}\sin \psi \sin \theta }{H_{11}\cos \psi \sin \theta +H_{22}\sin \psi
\cos \theta }
\end{equation*}
which has to be true whatever $\psi .$ Obviously, it is possible if and only
if $H_{11}=H_{22}.$ Using $\left( 35\right) ,$ we conclude that a NSC for a
bi-filter to induce a rotation of angle $\theta $ is summarized by 
\begin{equation}
\left\{ 
\begin{array}{l}
H_{21}=-H_{12}=H_{11}\tan \theta \\ 
H_{22}=H_{11} \\ 
A^{z}\left( t\right) =\frac{1}{\cos \theta }\mathcal{H}_{11}\left[ \mathbf{A}%
^{0}\right] \left( t\right)%
\end{array}
\right.
\end{equation}
except $\theta \neq \frac{\pi }{2}\func{mod}\pi .$ For this particular case,
we have 
\begin{equation*}
\left\{ 
\begin{array}{l}
H_{11}=H_{22}=0,\text{ \ }H_{12}=-H_{21} \\ 
A^{z}\left( t\right) =-\mathcal{H}_{12}\left[ \mathbf{A}^{0}\right] \left(
t\right) .%
\end{array}
\right.
\end{equation*}
To summarize, bi-filters verifying $\left( 36\right) $ define
transformations composed by a rotation of the direction of polarization,
associated to a LIF for the amplitude. The LIF has input $\mathbf{A}^{0},$
output $\mathbf{A}^{z}$ and complex gain $H_{11}/\cos \theta $ (for $\theta
\neq \frac{\pi }{2}\func{mod}\pi ).$

Then a \textquotedblleft pure rotation\textquotedblright\ which retains the
amplitude corresponds to the bi-filter 
\begin{equation*}
\left\{ 
\begin{array}{c}
H_{11}=H_{22}=\cos \theta \\ 
H_{21}=-H_{12}=\sin \theta%
\end{array}
\right.
\end{equation*}
which is not surprising.

\subsection{Appendix 3: Unpolarized wave}

1) If $s_{xy}^{z}=0$ whatever the system of coordinates, we have, when $%
\theta =\left( \text{Ox, Ox}^{\prime }\right) $%
\begin{equation}
\left\{ 
\begin{array}{l}
s_{x^{\prime }y^{\prime }}^{z}=\left( s_{y}^{z}-s_{x}^{z}\right) \sin \theta
\cos \theta =0 \\ 
s_{x^{\prime }}^{z}=s_{x}^{z}\cos ^{2}\theta +s_{y}^{z}\sin ^{2}\theta \\ 
s_{y^{\prime }}^{z}=s_{y}^{z}\cos ^{2}\theta +s_{x}^{z}\sin ^{2}\theta%
\end{array}
\right.
\end{equation}
whatever $\theta ,$ and then 
\begin{equation*}
s_{x^{\prime }}^{z}=s_{y^{\prime }}^{z}=s_{x}^{z}=s_{y}^{z}.
\end{equation*}
The definition which we have taken for the unpolarized beam holds on the
whole process $\overrightarrow{\mathbf{E}^{z}}$ and not only on the
two-dimensional random variable $\left( E_{x}^{z}\left( t\right)
,E_{y}^{z}\left( t\right) \right) .$ It takes into account all $\left(
E_{x}^{z}\left( t\right) ,E_{y}^{z}\left( t^{\prime }\right) \right) .$ It
is a strong difference.

As an example, take $E_{y}^{z}\left( t\right) =\mathcal{H}\left[ \mathbf{E}%
_{x}^{z}\right] \left( t\right) ,$ where $\mathcal{H}\left[ ..\right] $ is
the Hilbert tranform and $\mathbf{E}_{x}^{z}$ is a real process. It is
wellknown that this implies $\rho _{xy}^{z}=0,\rho _{x}^{z}=\rho _{y}^{z}$
and these equalities remain true in any coordinates systems (the equations $%
\left( 37\right) $ are verified for variances and covariances)$.$ Also 
\begin{equation*}
\left\{ 
\begin{array}{l}
s_{x}^{z}=s_{y}^{z}=s_{x^{\prime }}^{z}=s_{y^{\prime }}^{z} \\ 
s_{xy}^{z}=s_{x^{\prime }y^{\prime }}^{z}=-is_{x}^{z}\text{sign}%
\end{array}
\right.
\end{equation*}
where sign$\omega =1$ for $\omega >0$ and $-1$ for $\omega <0.$ Though $\rho
_{xy}^{z}=0$ in any system ($s_{x}^{z}\left( \omega \right) $ is even), the
r.v. $E_{x}^{z}\left( t\right) $ and $E_{y}^{z}\left( t^{\prime }\right) $
are linked for different $t,t^{\prime }$ because the cross-spectrum is
different of 0$.$

2) A compensator corresponds to a bi-filter such that (in Oxy) 
\begin{equation*}
\mathbf{H=}\left[ 
\begin{array}{cc}
e^{-i\omega \theta _{x}} & 0 \\ 
0 & e^{-i\omega \theta _{y}}%
\end{array}
\right]
\end{equation*}
$\theta _{x}$ and $\theta _{y}$ are the delays applied to $E_{x}^{0}$ and $%
E_{y}^{0}.$ If we assume $s_{xy}^{0}=0,$ the first formula of $\left(
7\right) $ implies that $s_{xy}^{z}=0.$ Then a compensator maintains the
property of unpolarization when the strong definition is used.

If we take $\mathbf{E}_{x}^{0}=\mathbf{A+B,\mathbf{E}_{y}^{0}=A-B\,\ }$where 
$\mathbf{B}$\textbf{\ }is the Hilbert tranform of $\mathbf{A}$ with real $%
\mathbf{A},$ we obtain (sgn$\omega =1$ for $\omega >0$ and $-1$ for $\omega
<0)$%
\begin{equation*}
s_{x}^{0}=s_{y}^{0}=s_{x}^{z}=s_{y}^{z}=2s_{A},s_{xy}^{0}=-2is_{A}\text{sgn}
\end{equation*}
which implies $\rho _{xy}^{0}=0$ because $s_{A}^{0}\left( \omega \right) $
is even. Then, the beam is unpolarized at O in the weak sense but not in the
strong sense. However from $\left( 7\right) $%
\begin{equation*}
\left\{ 
\begin{array}{c}
s_{xy}^{z}\left( \omega \right) =-2ie^{i\omega \left( \theta _{y}-\theta
_{x}\right) }s_{A}\left( \omega \right) \text{sgn}\omega \\ 
\rho _{xy}^{z}=4\int_{0}^{\infty }s_{A}\left( \omega \right) \sin \omega
\left( \theta _{y}-\theta _{x}\right) d\omega .%
\end{array}
\right.
\end{equation*}
Obviously we do not generally have $\rho _{xy}^{z}=0,$ which proves that a
compensator does not maintain the unpolarization in the weak definition,
except if the spectral support of $s_{A}\left( \omega \right) $ is small
enough around some $\omega _{0}$ and $\omega _{0}\left( \theta _{y}-\theta
_{x}\right) $ close to a multiple of $\pi .$

From $\left( 7\right) $ and with the strong definition, the unpolarization
is maintained from O to $z$ if and only if 
\begin{equation*}
H_{12}H_{22}^{\ast }+H_{11}H_{21}^{\ast }=0.
\end{equation*}

\subsection{Appendix 4: the Beer-Lambert law}

1) The beam state $\overrightarrow{\mathbf{E}^{z}}=\left( \mathbf{E}_{x}^{z},%
\mathbf{E}_{y}^{z}\right) $ at $z$ is the result of the bi-filtering of $%
\overrightarrow{\mathbf{E}^{0}}$ by the $H_{jk}^{0z}$ or the bi-filtering of 
$\overrightarrow{\mathbf{E}^{u}},u<z,$ by the $H_{jk}^{uz}$ (of course they
are function of $\omega $ but we can omit this variable)$.$ This leads to
the equations (using elementary properties of circuits and looking at figure
2) 
\begin{equation*}
\left\{ 
\begin{array}{c}
H_{11}^{0z}=H_{11}^{0u}H_{11}^{uz}+H_{21}^{0u}H_{12}^{uz} \\ 
H_{12}^{0z}=H_{12}^{0u}H_{11}^{uz}+H_{22}^{0u}H_{12}^{uz} \\ 
H_{21}^{0z}=H_{11}^{0u}H_{21}^{uz}+H_{21}^{0u}H_{22}^{uz} \\ 
H_{22}^{0z}=H_{12}^{0u}H_{21}^{uz}+H_{22}^{0u}H_{22}^{uz}%
\end{array}
\right.
\end{equation*}
which are simplified in (from $\left( 23\right) $ which translates the
homogeneity of the medium$)$%
\begin{equation}
\left\{ 
\begin{array}{c}
H_{11}^{z}=H_{11}^{u}H_{11}^{z-u}+H_{21}^{u}H_{12}^{z-u} \\ 
H_{12}^{z}=H_{12}^{u}H_{11}^{z-u}+H_{22}^{u}H_{12}^{z-u} \\ 
H_{21}^{z}=H_{11}^{u}H_{21}^{z-u}+H_{21}^{u}H_{22}^{z-u} \\ 
H_{22}^{z}=H_{12}^{u}H_{21}^{z-u}+H_{22}^{u}H_{22}^{z-u}%
\end{array}
\right.
\end{equation}
For instance , we write the first equation under the form 
\begin{equation*}
\frac{H_{11}^{z+a}-H_{11}^{z}}{a}=H_{11}^{z}\frac{H_{11}^{a}-1}{a}+H_{21}^{z}%
\frac{H_{12}^{a}}{a}.
\end{equation*}
If we assume the existence of derivatives $h_{jk}^{z}=\frac{\partial }{%
\partial z}H_{jk}^{z}$ we obtain 
\begin{equation}
\left\{ 
\begin{array}{l}
h_{11}^{z}=H_{11}^{z}h_{11}^{0}+H_{21}^{z}h_{12}^{0} \\ 
h_{12}^{z}=H_{12}^{z}h_{11}^{0}+H_{22}^{z}h_{12}^{0} \\ 
h_{21}^{z}=H_{11}^{z}h_{21}^{0}+H_{21}^{z}h_{22}^{0} \\ 
h_{22}^{z}=H_{12}^{z}h_{21}^{0}+H_{22}^{z}h_{22}^{0} \\ 
\text{with \ \ }h_{jk}^{z}=\frac{d}{dz}H_{jk}^{z}%
\end{array}
\right.
\end{equation}
which includes the (realistic) conditions 
\begin{equation}
\text{lim}_{z\rightarrow 0}H_{11}^{z}=\text{lim}_{z\rightarrow 0}H_{22}^{z}=1%
\text{ \ and \ \ lim}_{z\rightarrow 0}H_{12}^{z}=\text{lim}_{z\rightarrow
0}H_{21}^{z}=0.
\end{equation}
The system can be written as the matricial equation 
\begin{equation}
\mathbf{h}^{u}=\mathbf{PH}^{u},\text{ \ \ }\mathbf{P}=\left[ 
\begin{array}{cccc}
h_{11}^{0} & 0 & h_{12}^{0} & 0 \\ 
0 & h_{11}^{0} & 0 & h_{12}^{0} \\ 
h_{21}^{0} & 0 & h_{22}^{0} & 0 \\ 
0 & h_{21}^{0} & 0 & h_{22}^{0}%
\end{array}
\right]
\end{equation}
Three cases can be highlighted, following the properties of \textbf{P}. We
assume that $H_{jk}^{\infty }=0$ because a wave in a unlimited medium is
evanescent. This condition cancels constants which can appear in solutions
of the system.

2) We have three possibilities which are detailed below. 
\begin{equation*}
\text{Case }1:\left\{ 
\begin{array}{c}
\left( h_{11}^{0}-h_{22}^{0}\right) ^{2}+4h_{12}^{0}h_{21}^{0}\neq 0\text{
and }h_{11}^{0}h_{22}^{0}\neq h_{12}^{0}h_{21}^{0} \\ 
H_{jk}^{z}=c_{jk1}e^{\lambda _{1}z}+c_{jk2}e^{\lambda _{2}z}%
\end{array}
\right.
\end{equation*}
where $\lambda _{1},\lambda _{2}$ are distinct eigenvalues of \textbf{P (}%
which have negative real parts for a passive medium). The conditions $\left(
40\right) $ lead to 
\begin{equation}
\left\{ 
\begin{array}{l}
H_{11}^{z}=d_{11}e^{z\lambda _{1}}+\left( 1-d_{11}\right) e^{z\lambda _{2}}
\\ 
H_{12}^{z}=d_{12}e^{z\lambda _{1}}-d_{12}e^{z\lambda _{2}} \\ 
H_{21}^{z}=d_{21}e^{z\lambda _{1}}-d_{21}e^{z\lambda _{2}} \\ 
H_{22}^{z}=d_{22}e^{z\lambda _{1}}+\left( 1-d_{22}\right) e^{z\lambda _{2}}%
\end{array}
\right.
\end{equation}
with, using $\left( 39\right) $%
\begin{equation}
\begin{array}{cc}
d_{11}=\frac{h_{11}^{0}-\lambda _{2}}{\lambda _{1}-\lambda _{2}}, & d_{12}=%
\frac{h_{12}^{0}}{\lambda _{1}-\lambda _{2}} \\ 
d_{21}=\frac{h_{21}^{0}}{\lambda _{1}-\lambda _{2}}, & d_{22}=\frac{%
h_{22}^{0}-\lambda _{2}}{\lambda _{1}-\lambda _{2}}%
\end{array}%
\end{equation}
where $\lambda _{1}$ and $\lambda _{2}$ are (distinct, different of 0 and
with negative real parts) roots of 
\begin{equation}
\lambda ^{2}-\left( h_{11}^{0}+h_{22}^{0}\right) \lambda
+h_{11}^{0}h_{22}^{0}-h_{12}^{0}h_{21}^{0}=0
\end{equation}
Conversely, $\left( 42\right) $ with $\left( 43\right) $ verify $\left(
38\right) .$ Moreover formulae $\left( 11\right) $ imply the invariance of $%
\lambda _{1}$ and $\lambda _{2}$ in any rotation. We obtain the set of $%
K_{jk}^{z}$ fitted to Ox'y' by replacing $h_{jk}^{0}$ by $k_{jk}^{0}$=$\frac{%
\partial }{\partial z}K_{jk}^{0}$ in $\left( 42\right) ,\left( 43\right) $.
This case is developped in section 4.2.

\begin{equation*}
\text{Case }2:\left\{ 
\begin{array}{c}
\left( h_{11}^{0}-h_{22}^{0}\right) ^{2}+4h_{12}^{0}h_{21}^{0}\neq 0\text{
and }h_{11}^{0}h_{22}^{0}=h_{12}^{0}h_{21}^{0} \\ 
H_{jk}^{z}=c_{jk}e^{\lambda z}%
\end{array}
\right.
\end{equation*}
where $\lambda =h_{11}^{0}+h_{22}^{0}$ is the eigenvelue of \textbf{P}
assumed different of 0. Conditions $\left( 40\right) $ imply 
\begin{equation*}
c_{11}=c_{22}=1,\text{ \ }c_{12}=c_{21}=0.
\end{equation*}
Obviously the usual Beer-Lambert law is verified. Each component is weakened
and delayed through a quantity proportional to $z.$ 
\begin{equation*}
\text{Case }3:\left( h_{11}^{0}-h_{22}^{0}\right)
^{2}+4h_{12}^{0}h_{21}^{0}=0\text{ .}
\end{equation*}
In the last case, \textbf{P} has only one eigenvalue (of order 4) wich
corresponds to a proper subspace of dimension 2. Actually, we find same
results as in the case 2.

Then we have shown that the system $\left( 38\right) $ has an unique
solution most of the time depending of two parameters (for instance $%
h_{11}^{0},h_{12}^{0})$. However they may be functions of the frequency $%
\omega /2\pi .$ Consequently bi-filters are able to model propagation of
electromagnetic beams through continuous (and stationary) media such as free
space or optical fiber or coaxial cable.

\subsection{Appendix 5: a class of beams}

The simplest model of incoherent light is described by 
\begin{equation*}
\left\{ 
\begin{array}{c}
X\left( t\right) =\sum_{j}e^{i\omega _{0}\left( t-t_{j}\right) }h\left(
t-t_{j}\right) \cos \Theta _{j} \\ 
Y\left( t\right) =\sum_{j}e^{i\omega _{0}\left( t-t_{j}\right) }h\left(
t-t_{j}\right) \sin \Theta _{j}%
\end{array}
\right.
\end{equation*}
where \textbf{t}=$\left\{ t_{n},n\in \mathbb{Z}\right\} $ is an homogeneous
Poisson process with parameter $\lambda ,$ the $\Theta _{n}$ are random
variables independent of \textbf{t} and between them, and $h\left( t\right) $
is regular enough. Each term represents the emission by the particle $j$ at
the time $t_{j}$ in the direction $\Theta _{j}.$ Straigthforward
calculations yield (assuming $H\left( \omega _{0}\right) =0~$\ to suppress
some continuous component$)$%
\begin{equation*}
\left\{ 
\begin{array}{l}
s_{X}\left( \omega \right) =\frac{1}{4\pi }\left\vert H\left( \omega -\omega
_{0}\right) \right\vert ^{2}\text{E}\left[ \cos ^{2}\Theta \right] \\ 
s_{Y}\left( \omega \right) =\frac{1}{4\pi }\left\vert H\left( \omega -\omega
_{0}\right) \right\vert ^{2}\text{E}\left[ \sin ^{2}\Theta \right] \\ 
s_{XY}\left( \omega \right) =\frac{1}{4\pi }\left\vert H\left( \omega
-\omega _{0}\right) \right\vert ^{2}\text{E}\left[ \sin \Theta \cos \Theta %
\right] \\ 
H\left( \omega \right) =\int_{-\infty }^{\infty }h\left( u\right)
e^{-i\omega u}du%
\end{array}
\right.
\end{equation*}
Taking $\Theta _{n}$ uniformly distributed leads to (this means that the
elementary emitters have no favourite polarization) 
\begin{equation*}
\left\{ 
\begin{array}{l}
s_{X}\left( \omega \right) =s_{Y}\left( \omega \right) =\frac{1}{4\pi }%
\left\vert H\left( \omega -\omega _{0}\right) \right\vert ^{2} \\ 
s_{XY}\left( \omega \right) =0%
\end{array}
\right.
\end{equation*}
Then, the wave with components $\mathbf{X,Y}$ is unpolarized.

When Pr$\left[ \Theta =\theta \right] =1,$ we have a polarized beam in the
direction $\theta :$%
\begin{equation*}
\left\{ 
\begin{array}{l}
s_{X}\left( \omega \right) =\frac{1}{4\pi }\left\vert H\left( \omega -\omega
_{0}\right) \right\vert ^{2}\cos ^{2}\theta \\ 
s_{Y}\left( \omega \right) =\frac{1}{4\pi }\left\vert H\left( \omega -\omega
_{0}\right) \right\vert ^{2}\sin ^{2}\theta \\ 
s_{XY}\left( \omega \right) =\frac{1}{4\pi }\left\vert H\left( \omega
-\omega _{0}\right) \right\vert ^{2}\sin \theta \cos \theta%
\end{array}
\right.
\end{equation*}
Other laws for $\Theta $ give a large choice of situations in the form (for
real $\alpha ,\beta )$ 
\begin{equation*}
\left\{ 
\begin{array}{l}
s_{X}\left( \omega \right) =\alpha \left\vert f\left( \omega \right)
\right\vert ^{2} \\ 
s_{Y}\left( \omega \right) =\left( 1-\alpha \right) \left\vert f\left(
\omega \right) \right\vert ^{2} \\ 
s_{XY}\left( \omega \right) =\beta \left\vert f\left( \omega \right)
\right\vert ^{2}%
\end{array}
\right.
\end{equation*}
where $\left\vert \beta \right\vert \leq \sqrt{\alpha \left( 1-\alpha
\right) }.$ Conversely, we can find a probability law for each value of $%
\left( \alpha ,\beta \right) .$


\begin{thebibliography}{99}
\bibitem{Cram} H. Cramer, M. R. Leadbetter, \textit{Stationary and related
stochastic processes, }Wiley, New-York, 1967.

\bibitem{Ette} P. C. Etter, \textit{Underwater Acoustic modelling,}
Elsevier, Amsterdam, 1991.

\bibitem{Kuc} R. Kuc, \textit{Clinical application of an ultrasound
attenuation coefficient estimation technique for liver pathology
characterization, }IEEE Trans. Biomed. Eng., BME-27 (6) (1980) 312-319.

\bibitem{Laca1} B. Lacaze, \textit{A Theoretical Exposition of Stationary
Processes Sampling, }Samp. Th. Sign. Im. Proc. 4 (3) (2005) 201-230.

\bibitem{Laca2} B. Lacaze,\textit{\ Gaussian Delay Models for Light
Broadenings and Redshifts, }Elec. J. of Theoretical Physics, 6 (20) (2009)
385-398.

\bibitem{Laca3} B. Lacaze, \textit{Gaps of FSO beams with the Beer-Lambert
law, }Applied Optics, 48 (14) (2009) 2702-2706.

\bibitem{Laca4} B. Lacaze, \textit{Random Propagation Times in Ultrasonics,}
Waves in Random and Complex Media, (2010).

\bibitem{Laca5} B. Lacaze, \textit{About the Stokes decomposition theorem, }%
to appear.

\bibitem{Laca6} B. Lacaze, \textit{An unifying model for spectra of
transmitted monochromatic waves, }Wave Motion, 44 (2006) 70-76.

\bibitem{Lewi} P. A. Lewin, \textit{\textquotedblleft Quo vadis medical
ultrasound\textquotedblright , }Ultrasonics, 42 (4-2004) 1-5.

\bibitem{Luka} E. Lukacs, \textit{Characteristic Functions, }2$^{\text{nd}}$%
ed., Griffin, London, 1970.

\bibitem{Papo} A. Papoulis, \textit{Probability, Random Variables and
Stochastic Processes, }McGraw-Hill, 1965.

\bibitem{Park} K. J. Parker, R. M. Lerner, R. C. Waag, \textit{Attenuation
of Ultrasound: Magnitude and Frequency Dependence for Tissue
Characterization, }Radiology, 153 (1984) 785-788.

\bibitem{Paul} A. J. Paulraj, D. A. Gore, R. U. Nabar, H. Bolcskei, \textit{%
An Overview of MIMO Communications- A Key to Gigabit Wireless, }Proc. IEEE
92 (2) (2004) 198-218.

\bibitem{Mand} L. Mandel, E. Wolf, \textit{Optical Coherence and Quantum
Optics, }Cambridge Un. Press, 2nd Ed. 2008.

\bibitem{Szab} T. L. Szabo, \textit{Causal theories and data for acoustic
attenuation obeing a frequency power law, }J. Acoust. Soc. Amer. 97 (1)
(1-1995) 14-24.

\bibitem{Venk} R. Venkataramani, Y. Bresler, \textit{Sampling Theorems for
Uniform and Periodic Nonuniform MIMO Sampling of Multibands Signals, }IEEE
Trans. on SP 51 (12) (2003) 3152-3163.

\bibitem{Wolf} M. Born, E. Wolf,\textit{\ Principles of Optics, }5th ed.
Pergamon Press 1975.

\bibitem{Wolf1} E. Wolf, \textit{Can a light be considered to be the sum of
a completely polarized and a completely unpolarized beam?}, Optics Letters
33 (7) (4-2008).
\end{thebibliography}
\end{document}